\documentclass[twocolumn]{article}

\usepackage[utf8]{inputenc}
\usepackage{microtype}
\usepackage[numbers,square]{natbib}
\usepackage{booktabs}
\usepackage{color}
\usepackage{url}
\usepackage{makecell} 
\usepackage{enumitem}
\usepackage{graphicx}
\usepackage{float}

\usepackage{soul}

\usepackage{authblk}

\usepackage{dblfloatfix} 

\usepackage{subcaption}

\newcommand\numAdditionalListeners{8 }

\title{Investigating the impact of stereo processing -- a study for extending the Open Dataset of Audio Quality (ODAQ)}

\author[1]{Sascha Dick}
\author[2]{Christoph Thompson}
\author[3]{Chih-Wei Wu}
\author[1]{Pablo Delgado}
\author[3]{Phillip A. Williams}
\author[1*]{Matteo Torcoli}

\affil[1]{Fraunhofer Institute for Integrated Circuits IIS, Erlangen, Germany}
\affil[2]{Ball State University, Muncie, USA}
\affil[3]{Netflix, Inc., Los Gatos, USA}

\usepackage[pscoord]{eso-pic}
\newcommand{\placetextbox}[3]{
\setbox0=\hbox{#3}
\AddToShipoutPictureFG*{
\put(\LenToUnit{#1\paperwidth},\LenToUnit{#2\paperheight}){\vtop{{\null}\makebox[0pt][c]{#3}}}%
}%
}%
\date{}

\begin{document}

\twocolumn[
\maketitle

\begin{abstract}
In this paper, we present an initial study for extending Open Dataset of Audio Quality (ODAQ) towards the impact of stereo processing. Monaural artifacts from ODAQ were adapted in combinations with left-right (LR) and mid-side (MS) stereo processing, across stimuli including solo instruments, typical wide stereo mixes and and hard-panned mixes. Listening tests in different presentation context -- with and without direct comparison of MS and LR conditions -- were conducted to collect subjective data beyond monaural artifacts while also scrutinizing the listening test methodology. The ODAQ dataset is extended with  new material along with subjective scores from 16 expert listeners. The listening test results show substantial influences of the stimuli's spatial characteristics as well as the presentation context. Notably, several significant disparities between LR and MS only occur when presented in direct comparison. The findings suggest that listeners primarily assess timbral impairments when spatial characteristics are consistent and focus on stereo image only when timbral quality is similar. The rating of an additional mono anchor was overall consistent across different stereo characteristics, averaging at 65 on the MUSHRA scale, further corroborating that listeners prioritize timbral over spatial impressions.
\end{abstract}
\vspace{12pt}
]

\placetextbox{0.5}{0.08}{\fbox{\parbox{\dimexpr\textwidth-2\fboxsep-2\fboxrule\relax}{\footnotesize \centering Accepted for presentation at the Audio Engineering Society (AES) 159th Convention, October 2025, Long Beach, USA.}}}
\section{Introduction}

{
  \renewcommand{\thefootnote}{\fnsymbol{footnote}}
  \footnotetext[1]{M. Torcoli was affiliated with Fraunhofer IIS at the time the work was conducted and is now with Amplifon.}
}

Investigating the perceptual impact of different signal impairments and gathering subjective ground truth data are instrumental in the development of objective quality metrics and optimization of audio processing methods. However, openly available datasets are scarce, especially when it comes to the influence of stereo and multichannel signal processing. To address the scarcity of openly available datasets, we have recently presented the Open Dataset of Audio Quality (ODAQ) \cite{Torcoli2024ODAQ,dick2024ODAQ} consisting of audio material with different signal impairments and associated subjective scores, with a primary focus on monaural artifacts and changes in timbre. 

This paper presents a study extending ODAQ with new data on the impact of stereo processing and binaural hearing on perceived quality. 
The impact of the spatial image on overall quality perception has been investigated in the literature \cite{rumsey2005relative, flessner2019subjective, qiao2022prediction, conetta2015spatial1, conetta2015spatial2}, and it has been found that overall quality is dominated by timbral quality. According to Rumsey et. al \cite{rumsey2005relative}, a regression analysis shows that the spatial quality accounts for approximately \ 30\% of the basic audio quality. 

To objectively assess spatial quality, recent efforts have focused on data-driven approaches using synthetic data \cite{panah2025binaqual} or augmented data \cite{Zheng2025HAPGSAQUAM}. However, the associated datasets are typically not made openly available, increasing the difficulty in reproducing and validating these methods. 

\begin{figure}[t]
    \centering
    \begin{subfigure}{\linewidth}
        \includegraphics[width=0.95\linewidth]{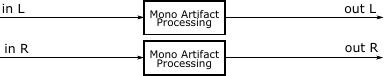}
        \caption{{\label{fig:msprocessing:LR}}Left-Right (LR) Processing}
    \end{subfigure}

    \vspace{0.25cm}
    
    \begin{subfigure}{\linewidth}
        \includegraphics[width=0.95\linewidth]{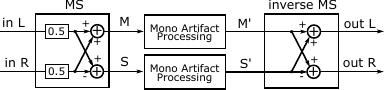}
        \caption{{\label{fig:msprocessing:MS}}Mid-Side (MS) Processing}
    \end{subfigure}
    \caption{{\label{fig:msprocessing}}Stereo processing techniques in our test signal generation pipeline}
\end{figure}

In a previous study \cite{dick2017}, we evaluated the impact of various types of isolated signal degradations on perceived quality using subjective listening tests. The focus was on monaural signal degradations, and the respective processing was applied independently to the left/right (LR) stereo channels as illustrated in Figure~\ref{fig:msprocessing:LR}. This, however, subsequently affected the stereo image according to the reports from the listeners.
In audio coding, Mid-Side (MS)-Stereo  \cite{herre1992ms,johnston1992sum} is widely used for joint coding of stereo signals to exploit redundancy as well as to account for perceptual properties of binaural hearing.
To minimize alterations of stereo image in the first iteration of ODAQ \cite{Torcoli2024ODAQ}, we employed MS-Stereo pre-/post-processing as illustrated in Figure~\ref{fig:msprocessing:MS} for artifact types that are prone to spatial artifacts. 

Besides the actual signal impairments, also the \emph{presentation context} -- i.e., the multitude of different conditions and quality levels presented together in a test  -- can have substantial impact on subjective ratings and scale usage, as reported by Zielinsky et al. \cite{zielinski:2008}. For instance, smaller differences between two conditions may only be noticeably in a direct comparison, but not when individually presented in different contexts.

To systematically study the impact of commonly used stereo processing techniques while addressing the data scarcity issue, this paper builds on our previous investigation and focuses on the comparison of LR and MS stereo processing approaches within a cohesive listening experiment. 
It also investigates the influence of the presentation context, whether a direct comparison of LR and MS conditions is available or not to the listeners.
The processed material and subjective scores are available online in the ODAQ dataset\footnote{\url{https://github.com/Fraunhofer-IIS/ODAQ/}\label{first_footnote}}
\footnote{\url{https://doi.org/10.5281/zenodo.17162670}\label{second_footnote}}.

\section{Methodology}
\subsection{Overview}
A challenge in designing listening tests related to stereo processing is the impact on both spatial and timbral characteristics, as well as their corresponding contributions to the basic audio quality. As a step-by-step approach, this study investigates the commonly-used stereo processing techniques under the influence of different presentation context and anchor conditions. 

\subsection{Simulated Coding Artifacts}

New test signals were generated following the methods used in ODAQ \cite{Torcoli2024ODAQ} and originally presented in \cite{dick2017} to create two isolated types of audio coding artifacts by forcing audio coders into controlled, sub-optimal operation modes. The following two signal impairments were chosen as they primarily affect timbre were found to be prone to interact with stereo coding in our previous studies:
\begin{enumerate}[itemsep=0pt,topsep=0pt, leftmargin=*]
    \item \textbf{Quantization Noise (QN)} simulated by additive noise with a frequency-dependent noise-to-mask ratio (NMR)
    \item \textbf{Spectral Holes (SH)} due to coarse quantization, here, simulated through randomly inserted spectral holes with fixed, controlled probability.
\end{enumerate}

The QN impairments were used in the previous experiments \cite{dick2017,Torcoli2024ODAQ} primarily in the context of pre-echoes (PE), i.e.\ temporal smearing of quantization noise around transients. Here, it is employed to assess the impact of additive quantization noise for general signals. This is also motivated by the Binaural Masking Level Difference (BMLD) experiments in literature \cite{BlauertBook,Robinson63} using tones and noise in different interaural correlation conditions.

Both artifact types originate from quantization in the frequency domain, but represent perceptually different effects. Coarse quantization of signal components to zero can result in spectral holes or islands that fluctuate over time and are perceived as additional signal components (also called ``birdies''), as simulated by SH. As long as the quantization maintains a coarser representation of the original signal, it results in additional quantization noise as simulated by QN. Such quantization noise roughly follows the signal's temporal and spectral envelope and is thus perceived as a noise-like distortion of the signal. 
In the previous experiments, the quantization noise for PE was generated by applying actual quantization at appropriate levels. However, at low NMR, this can lead to SH-like artifacts. To keep the two artifact types under test clearly distinguished, QN is generated here by adding random noise of the respective level to the test signal rather than applying quantization.

The extension of the above-mentioned monaural artifacts to stereo signals was achieved via two different strategies, namely the LR and MS processing techniques as shown in Figure~\ref{fig:msprocessing}. Thus, we generated two stereo variants for both SH and QN, each in five quality levels. 
The detailed parameters for the generation of the quality levels are given in Table~\ref{tbl:artifacts}.

\begin{table}[!t]
    \begin{footnotesize}
    \resizebox{\columnwidth}{!}{%
    \centering
    \setlength{\tabcolsep}{4pt}
    \begin{tabular}{c c c c c c c}
        \toprule
        \textbf{Method} & \textbf{Parameter} & \multicolumn{5}{c}{\textbf{Quality Level}} \\[2pt] \cmidrule{3-7}
        \textbf{} & \textbf{} & \textbf{Q1} & \textbf{Q2} & \textbf{Q3} & \textbf{Q4} & \textbf{Q5}\\ \midrule
        QN & NMR [dB]& 0 & 6 & 12 & 18 & 24 \\
        SH & hole prob. [\%] & $70$ & $50$ & $30$ & $20$ & $10$ \\ 
        \bottomrule
    \end{tabular}
    }
    \captionof{table}{\label{tbl:artifacts} Monaural artifacts and quality levels.}
    \label{TableArtifactParameters}
    \end{footnotesize}
\end{table}

\subsection{Audio Material}
Test signals were chosen to encompass a wide range of spatial characteristics in three levels of stereo width: recordings of solo instruments in the center stereo image; wider stereo mixes of music arrangements; and artificially mixed items with hard-panned stereo positions. For each artifact type, two specific items (i.e., solo instrument and music mix) were chosen, and two hard-panned items were tested for both artifact types, resulting in four items per artifact and a total of six different signals. The format of the test samples is 48\,kHz / 24\,bit stereo, and the duration is between 10\,s and 16\,s.
A detailed description of the test items is given in Table~\ref{tbl:items}.
The items \emph{violin}, \emph{glock} and \emph{Pop} are taken directly from the existing ODAQ dataset, and the \emph{panDialogF} and \emph{panDialogM} are remixes of items from the Dialogue Enhancement (DE) subset in ODAQ using the available Foreground (FG) and Background (BG) stems. 
Another music mix with wider stereo image \emph{RnB} has been added to the dataset for this experiment.  

\begin{table}[!t]
    \begin{footnotesize}
    \resizebox{\columnwidth}{!}{
    \centering
    \setlength{\tabcolsep}{4pt}
    \begin{tabular}{l l c}
        \toprule
        \textbf{Item} & \textbf{Description} & \textbf{Stereo} \\ \midrule
        violin & Solo violin playing mid-tempo melody & center \\ 
        glock & Solo glockenspiel playing distinct notes & center\\ 
        Pop & Upbeat pop music & wide \\
        RnB & Percussive RnB music with male vocals & wide\\
        panDialogM & Male speech + ambient music & hard pan\\
        panDialogF & Female speech + percussive music & hard pan \\ 
        \bottomrule
    \end{tabular}
    }
    \captionof{table}{\label{tbl:items} List of test stimuli.}
    \end{footnotesize}
\end{table}

\subsection{Experimental Setup}
A listening test following MUSHRA \cite{MUSHRA} methodology was conducted to evaluate subjective quality of the different processing methods and to also explore the influence of presentation context. 
To investigate the impact of the stereo characteristics, we presented the conditions in different presentation context, with and without direct comparison between LR and MS.

The basic set of \emph{separated trials} consists of a total of four combinations of the two stereo variants and the two artifact types, each presented at five different quality levels. These trials are denoted here as \emph{SHLR}, \emph{SHMS}, \emph{QNLR}, and \emph{QNMS}. They allow a direct comparison of all quality levels of one processing type, but only for indirect comparison of MS and LR at each quality level.

In addition, we introduced another set of \emph{mixed trials} that includes MS and LR items of same quality level and artifact type within the same trial for direct comparison to study the significance of presentation context, denoted as \emph{SHmix} and \emph{QNmix}. 
To keep the number of conditions under test reasonable, we selected two quality levels per artifact for this direct comparison of stereo processing methods. The selection of the respective trial configurations is also indicated in Table~\ref{tbl:trials}.
\begin{table}[!t]
    \begin{footnotesize}
    \resizebox{\columnwidth}{!}{%
    \centering
    \setlength{\tabcolsep}{3pt} 
    \renewcommand{\arraystretch}{1.2} 
    \begin{tabular}{c c c  c c c c  c c  c c}
        \toprule
        \textbf{Quality Level} & \multicolumn{2}{c}{\textbf{Q1}} & \multicolumn{2}{c}{\textbf{Q2}} & \multicolumn{2}{c}{\textbf{Q3}} & \multicolumn{2}{c}{\textbf{Q4}} & \multicolumn{2}{c}{\textbf{Q5}} \\
        \textbf{Stereo Proc.} & LR & MS & LR & MS & LR & MS & LR & MS & LR & MS \\
        \cmidrule{1-1}
        \textbf{Trial}\\
        \midrule
        SHLR   & x & - & x & - & x & - & x & - & x & - \\
        QNLR   & x & - & x & - & x & - & x & - & x & - \\
        SHMS   & - & x & - & x & - & x & - & x & - & x \\
        QNMS   & - & x & - & x & - & x & - & x & - & x \\
        SHmix  & - & - & - & - & x & x & - & - & x & x \\
        QNmix  & - & - & x & x & x & x & - & - & - & - \\
        \bottomrule
    \end{tabular}
    }
    \captionof{table}{\label{tbl:trials} Stereo processing methods and quality levels for selected trial series under test.}
    \label{TableArtifactParameters}
    \end{footnotesize}
\end{table}

For comparability to previous results, a small overlap with the conditions in the original ODAQ dataset was included in the test set (SHMS conditions for \emph{Pop} and \emph{glock} correspond to SH in \cite{Torcoli2024ODAQ}).

All tests included the MUSHRA standard $3.5$\,kHz and $7.0$\,kHz lowpass anchors.
To further assess the contribution of spatial characteristics to the basic audio quality, a mono anchor was included in the SHmix and QNmix trials. 
All the trials have the same number of conditions under test.
The complete test consisted of 22 trials/screens, and each trial contained seven test conditions and a hidden reference.

\subsection{Test Subjects and Listening Environment}

The subjective listening test was conducted at Ball State University (BSU) in the US. 
The test subjects were undergraduate students enrolled in the Music Media Production program in the School of Music. These students are trained musicians and receive musical ear training in addition to the technology courses in the curriculum.
All listening tests were conducted in acoustically damped listening rooms that are suitable for audio mixing or screening. For playback, Beyerdynamic DT770 Pro 250 Ohm closed-back headphones were used in combination with a professional 24-bit soundcard.
Before the actual test, each listener participated in a brief training session consisting of two items, to familiarize themselves with the methodology, user interface, and test conditions. The training items are not included in the actual test. 

The tests were completed by a total of 16 subjects with a mean age of 20.8 years (SD 0.9 years), with an average of 3.1 years of experience (SD 1.3 years).
Additional test results were obtained for a smaller set of \numAdditionalListeners expert listeners for conditions that the main test revealed to be of further interest (hard-panned items in both SHMS and SHLR conditions). Unless indicated otherwise, the results presented in the following are based on the main set of 16 listeners.  Results from the smaller listener group are marked in plots by grey X.

\section{Results and Discussion}
The results of the listening tests are illustrated with focus on different aspects in Figures \ref{fig:results:overallQ} to \ref{fig:results:itemwise}, showing mean scores and bootstrapped 95\% confidence intervals. 
For consistent visual distinction across all figures, LR conditions are indicated by empty markers, while MS conditions use filled markers. Furthermore, the QN artifacts are indicated by cool colors (blue-green) and the SH artifacts by warm colors (orange-red).

\subsection{Overall Results and Quality Scale Coverage}
\begin{figure}[tb]
\centering
\includegraphics[width=0.9\linewidth]{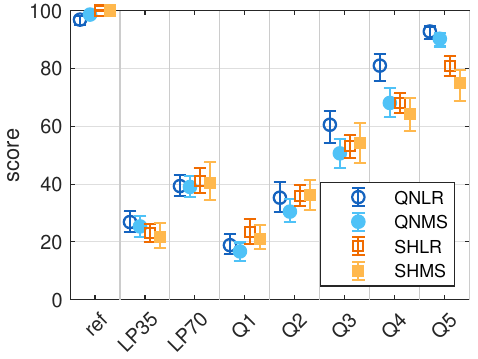}
\caption{{\label{fig:results:overallQ}}Overall results per processing method and quality level. (mean and 95\% CI, 16 participants) The results show that a good coverage of the quality scale was achieved.}
\end{figure}

Figure~\ref{fig:results:overallQ} shows the overall results for the five quality levels under test and the anchors in the separated trials. As it can be seen, a good coverage of the quality scale with relatively uniform spacing was achieved with the chosen quality levels. However, the quality levels for QN reach further into the upper range of the scale, whereas for SH the coverage of the quality scale is more compressed between 20 and 80 points. 
The chosen maximum quality parameters for the generation of QN with NRM at 24\,dB reach a higher quality range than the maximum quality for SH with 10\% hole probability. This suggests a parametrization of SH aiming for higher quality should be considered for future experiments. 

\subsection{Consistency with Previous Results}
\begin{figure}[htb]
\includegraphics[width=\linewidth]{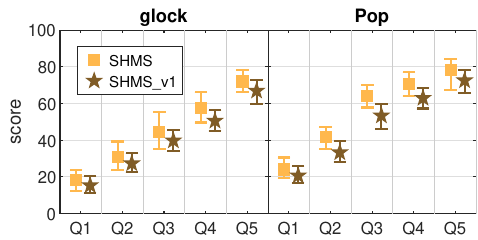}
\caption{{\label{fig:results:itemwiseSHvsV1}}Results for SHMS compared to corresponding condition from previous test as `SHMS\_v1' \cite{dick2024ODAQ}, showing good agreement.
}
\end{figure}

An overlap of some conditions with our previously presented results in \cite{dick2024ODAQ} was included to test consistency. Figure~\ref{fig:results:itemwiseSHvsV1} shows the results for the corresponding items and quality levels.
There is a tendency for slightly higher average ratings in the present results, but overall, the results are in good agreement and confirm the consistency between listener pools.

\begin{figure*}[htb]
\centering
\includegraphics[width=\linewidth]{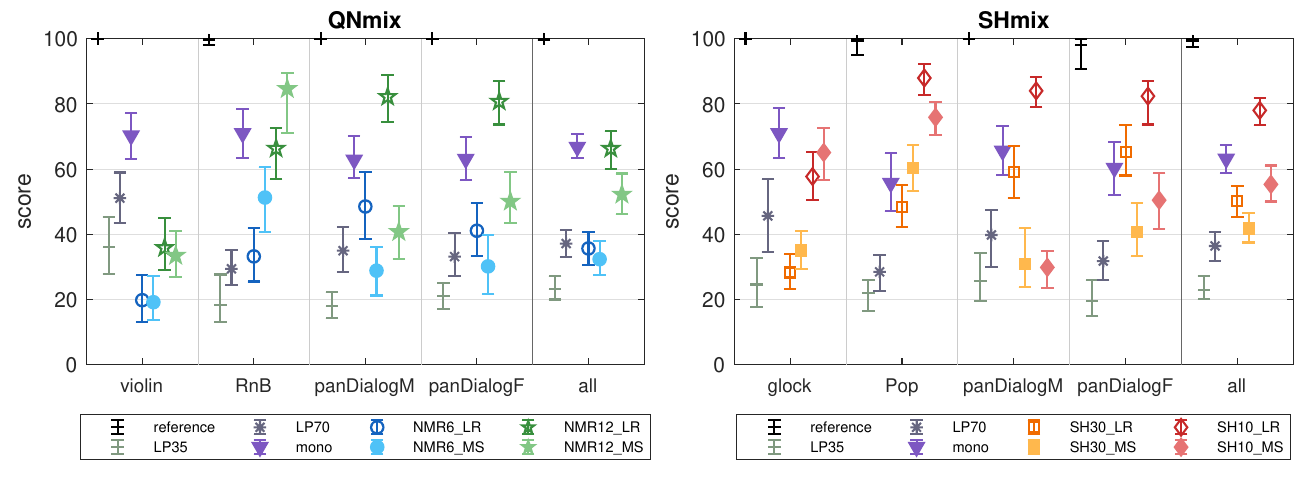}
\caption{{\label{fig:results:detailedLRMS}}Detailed results for mixed LR/MS tests (16 participants) showing mean scores and 95\% confidence intervals (CI) per experiment and item. The results show that preference for LR or MS coding is highly dependent on the item characteristic. Also, the mono anchor is rated remarkably high and consistent, even for items with hard-panned left-right stereo characteristics.}
\end{figure*}

\subsection{Mixed Tests: QNmix, SHmix}
Figure \ref{fig:results:detailedLRMS} shows the detailed results for QNmix and SHmix, i.e.,\ the tests with direct comparison of the LR vs.\ MS conditions at the same quality levels, for all individual items and conditions under test. The results show that the absolute rating of the conditions and especially the listener's preference for either the LR or MS conditions is highly item dependent.

\subsubsection{Direct Comparison of LR vs.\ MS}
For the hard-panned items (\emph{panDialogM}, \emph{panDialogF}), the LR conditions are consistently rated higher than the MS conditions, for both QN and SH artifacts. This is expected, as the joint coding of the distinct channels leads to audible interaction and crosstalk between the channels. 
For this type of item, the artificial generation of the signal distortions in the MS-domain results in audible interaction and crosstalk between the channels that goes beyond the more typical "spatial unmasking" properties that would be found in regular audio codecs. 

For the solo instrument items (\emph{violin} for QN and \emph{glock} for SH) the overall rating of the quality levels is lowest and relatively similar for LR and MS. (For SH, the mean rating of SHMS is slightly higher but not at significance level.) The results suggest that for those pure tonal items the tonal quality distortions outweigh any differences in the spatial characteristics.

For the stereo music mixes, in \emph{RnB} for QNmix for both quality levels and in \emph{Pop} for the lower quality level (SH30), the MS condition is rated significantly higher than LR. This is expected, as the MS coding better follows the spatial properties of those wide, but not discretely stereo-panned mixes. However, for SHmix at the higher quality level (SH10), the rating is reversed, with LR rated significantly above MS. Considering the signal characteristics, this is unexpected and should be considered for future investigations. A possible reason is that at the high quality level, the spatial fluctuation due to the rather sparse separate spectral holes in the left and right channel are less noticeable than the overall larger loss of energy that spectral holes in the M signal evoke in both ears. 

\subsubsection{Mono Anchor Condition}
A mono downmix was included to assess the influence of the overall spatial image and suitability of a mono signal as an anchor condition. 
On average, the mono condition was rated at 65 points, i.e.,\ well within the ``good'' range of the MUSHRA scale. The scores are also rather consistent over all items. 

The remarkably high rating of the mono anchor was slightly unexpected, especially for the hard-panned items where the distinct spatial distribution of sound sources is entirely lost. 
Arguably, the timbre of this condition is not affected except for potential slight comb-filter effects. This indicates that listeners overall put substantially more emphasis on the timbral quality than on the spatial properties. In that sense, the results are in line with the findings by \cite{rumsey2005relative}, suggesting that the spatial quality makes up for approximately 30\% of the basic audio quality. However, it is interesting to see that these findings that were obtained in a regression analysis seem to translate relatively well to approximately 30\% of the MUSHRA scale in our results. 

Furthermore, the rating of the mono condition is rather consistent across the very different stereo image characteristics of the different items. Consequently, it is well suited to serve as a stable anchor conditions in listening tests, e.g.\ in future experiments with focus on investigating spatial properties. 

\begin{figure*}[h!]
    \centering
    \begin{subfigure}{\textwidth}
        \includegraphics[width=\linewidth]{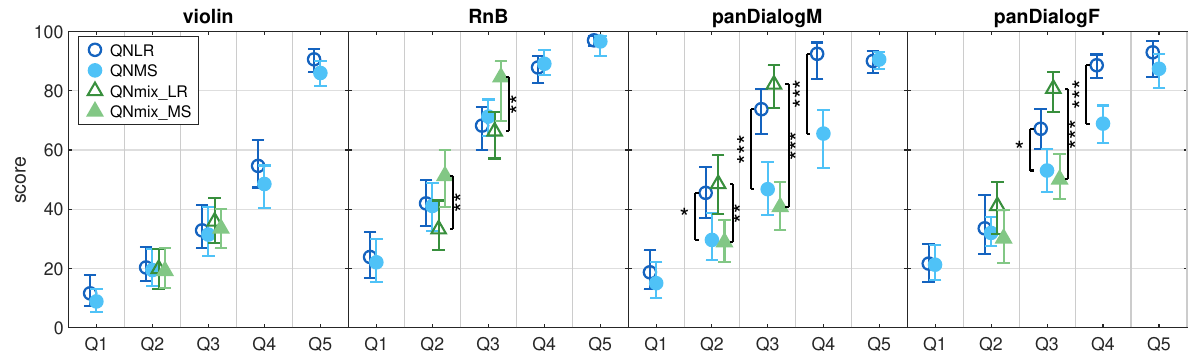}
        \caption{{\label{fig:results:itemwiseQN}}Quantization noise (QN) tests: QNLR, QNMS, QNmix.}
    \end{subfigure}
    \begin{subfigure}{\textwidth}
        \includegraphics[width=\linewidth]{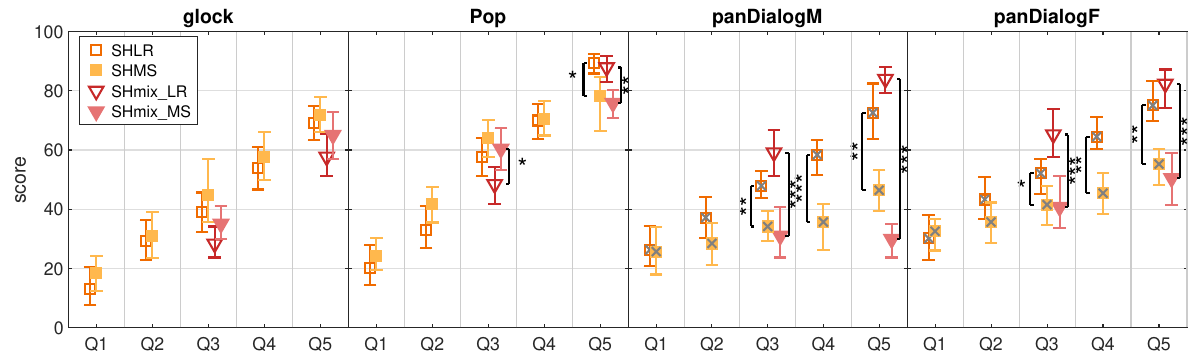}
        \caption{{\label{fig:results:itemwiseSH}}Spectral holes (SH) tests: SHLR, SHMS, SHmix}
    \end{subfigure}
    \caption{{\label{fig:results:itemwise}}Detailed results per item and quality level, combining separated and direct comparison of LR and MS conditions (mean and 95\% CI, 16 participants, results from additional \numAdditionalListeners participants marked by gray X). Significant differences between LR and MS conditions indicated in black (* p<0.5,** p<0.1, *** p<0.001). The results show several instances where the differences between LR and MS are only significant when listeners could directly compare them in presentation context, indicating that listeners primarily rate based on timbre quality and only put secondary focus on spatial properties.}
\end{figure*}

\subsection{Combined Results}

Figures \ref{fig:results:itemwiseQN} and \ref{fig:results:itemwiseSH} show the detailed scores for all the quality levels and the available LR and MS conditions per item, from the separated and mixed test variants together.
To focus on the comparison between LR and MS, significant differences between LR and MS conditions at the same quality level and presentation context type are indicated by black bars and asterisk markers to indicate significance levels. 

Overall, the results show that the differences in rating of MS and LR are more pronounced for the QNMix and SHmix trials where the listeners could directly compare between LR and MS. For multiple items the difference between LR and MS at a given quality level is negligible in the separated trials, but significantly different in direct comparison. 

For the pure instrumental items (violin, glock), no significant differences between MS and LR were found for any quality level or presentation context. This suggests that for these very tonal solo instrument signals, the overall quality is dominated by timbral artifacts.

For the regular stereo music mixes (RnB and Pop), the results show several instances where the separated trials have similar scores for LR and MS, but significantly different ratings in the mixed trials with direct comparison. (QNmix for RnB at Q2 and Q3; SHmix for Pop at Q3). This suggests that the direct comparison enables listeners to evaluate finer differences in the spatial characteristics that they cannot distinguish otherwise. It should also be noted that the ratings of a higher number of quality levels with similar artifact characteristics is likely also influenced by a spacing bias as reported by Zielinsky \cite{zielinski:2008}, i.e., the effect that listeners tend towards more uniform spacing across the entire quality scale for clearly distinguishable quality levels.

For the hard-panned items (panDialogM, panDialogF), any significant differences between LR and MS are consistently present for both the separated trials and corresponding mixed trials (where available in the results). For these extreme stereo characteristics, the differences between MS and LR appear to be obvious enough that listeners can rate them also without direct comparison. This indicates that listeners distinguish sufficiently larger differences in spatial characteristics independent of the presentation context.

Especially noteworthy differences between LR and MS are found for SH in panDialogM. Here, the MS condition has overall very low quality and for SHmix, the difference between LR and MS is over 50 points. 
A possible explanation is that for hard-panned items, holes in the `M' or `S' signal impede the inverse MS reconstruction and result in severe crosstalk between the left and right channel. For lower hole probability, this still can result in clearly audible blips and birdies. 
These findings suggest that a very strong impact on the stereo image can also result in cases where spatial properties are a predominant factor. However, the effect can also be interpreted as an individually strong timbral distortion to the quieter channel. 

In conclusion, the results indicate that for typical items without extreme spatial characteristics, listeners rate predominantly based on timbral quality. Only when different stereo characteristics are presented in a context that allows for direct comparison, listeners put additional attention to spatial properties. 

\section{Summary and Conclusion}
Listening tests were conducted to investigate the influence of stereo processing, spatial characteristics and presentation context on subjective quality.

The subjective results show that the preference for LR or MS processing is substantially influenced by the stereo characteristics of the test signal, with MS favored for recordings with a narrow stereo image and LR preferred for items with distinct stereo positioning. For signals with strong tonal properties, no significant impact of stereo processing was found, suggesting that obvious changes in timbral quality is the predominant factor. However, the results show that differences in ratings between LR and MS are significantly affected by the presentation context, with some conditions showing disparities only when direct comparison is enabled (e.g., SH-mix and QN-mix). It is hypothesized that listeners primarily assess timbral impairments when spatial characteristics are consistent, and focus on stereo image distortions only when timbral quality levels are similar; hence, stereo artifacts may serve as a "tie-breaker" in such scenarios.

The included mono anchor was consistently rated across items, which renders it a viable anchor of the quality scale. The relatively high ratings in the "good" range of the MUSHRA scale even for hard-panned stereo items are intriguing, but in line with findings in literature on the prioritization of timbral over spatial differences in subjective quality assessment .

In conclusion, the experimental results offer insights into rating of timbral and spatial impressions, designing listening tests for spatial audio quality, and emphasize the importance of presentation context. Furthermore, the addition of the test signals and subjective results to the ODAQ dataset mark a first step in expanding ODAQ towards stereo and multichannel audio.

\section{Acknowledgements}
We would like to thank Mhd Modar Halimeh, William Wolcott, and Emanu\"el Habets for their valuable support in creating ODAQ and reviewing this paper. We also warmly thank all participants in the listening tests.

\bibliographystyle{unsrt}
\begin{small}
\bibliography{refs}
\end{small}

\end{document}